# Delta-Sigma Modulator based Compact Sensor Signal Acquisition Front-end System

Joydeep Basu and Pradip Mandal

***Abstract*—The proposed delta-sigma modulator (ΔΣM) based signal acquisition architecture uses a differential difference amplifier (DDA) customized for dual purpose roles, namely as instrumentation amplifier and as integrator of ΔΣM. The DDA also provides balanced high input impedance for signal from sensors. Further, programmable input amplification is obtained by adjustment of ΔΣM feedback voltage. Implementation of other functionalities, such as filtering and digitization have also been incorporated. At circuit level, a difference of transconductance of DDA input pairs has been proposed to reduce the effect of input resistor thermal noise of front-end R-C integrator of the ΔΣM. Besides, chopping has been used for minimizing effect of Flicker noise. The resulting architecture is an aggregation of functions of entire signal acquisition system within the single block of ΔΣM, and is useful for a multitude of dc-to-medium frequency sensing and similar applications that require high precision at reduced size and power. An implementation of this in 0.18-$\mu$m CMOS process has been presented, yielding a simulated peak signal-to-noise ratio of 80 dB and dynamic range of 109dBFS in an input signal band of 1 kHz while consuming 100 $\mu$W of power; with the measured signal-to-noise ratio being lower by about 9 dB.**

***Index Terms*—Analog front-end, analog to digital conversion, ADC, delta-sigma modulation, instrumentation amplifier, sensor interface, signal acquisition, signal conditioning.**

## I. Introduction

THE analog electrical signal obtained from a sensor (e.g., accelerometer, thermocouple, biosensor, gas sensor, etc.) is usually very weak in magnitude, and is required to be conditioned for improving its quality and making it suitable for feeding to an analog-to-digital converter (ADC) [1], [2]. The task of signal conditioning that includes amplification and filtering, and subsequent task of digitization, are performed by signal acquisition front-end (or, readout) system that acts as an interface between the sensor and the digital processor [2], [3]. Although the performance and complexity of such a front-end system is principally determined by its application, a typical architecture (Fig. 1) consists of an *instrumentation amplifier* (IA) capable of providing sufficient amplification to the weak input signal from sensors, a *low-pass filter* (LPF) to remove high frequency noise components, and a *programmable gain amplifier* (PGA) stage for getting maximum output signal swing for input signals of varying strengths. Subsequently, the signal is passed through an *anti-aliasing filter* (AAF) and then digitized using ADC. The increasing popularity of portable sensor based devices demand compact analog front-end (AFE) with high power efficiency.

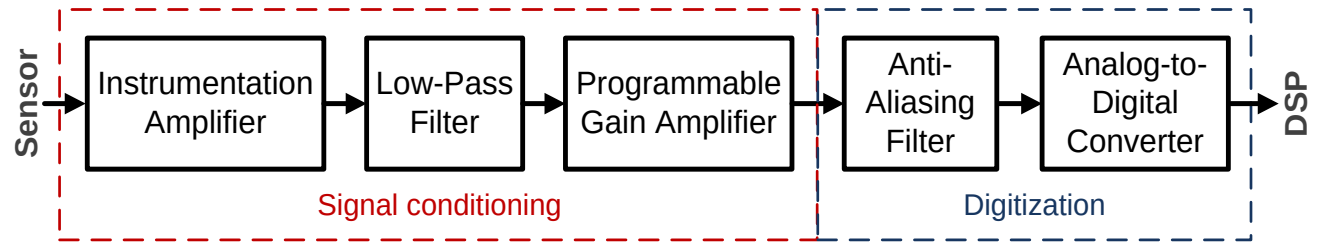

Fig. 1. Typical sensor signal acquisition system architecture.

A popular ADC type is *Delta-Sigma* (ΔΣ) ADC suitable for low and medium frequency applications where accuracy is more significant than speed [3], [4]. ΔΣ ADC consists of two consecutive blocks, namely an analog *ΔΣ modulator* (ΔΣM) followed by a digital *decimation filter*. The ΔΣM is made up of a loop-filter (low-pass filter (LPF), or integrator) and a coarse quantizer present in a negative feedback loop [4]–[6].

Reported AFE architectures have clear demarcation among the ADC and preceding signal conditioning section. Thus, the usual front-end becomes a long chain of successive circuit blocks performing different functionalities (as illustrated in Fig. 1). A closer inspection reveals that some blocks used in the chain have similar function, like the low-pass filtering function of the IA can be utilized as the LPF required in the subsequent ΔΣM; or equivalently, the input LPF stage of the modulator can be adapted to leverage the function of an IA as well. Hence, a judicious merging of the blocks/functions of the ΔΣM and signal conditioner seems feasible, leading towards abridged hardware requirement. This is the basis of the present work that puts forward an integrated second-order ΔΣM based precision signal acquisition architecture having potential for lower area and power requirement. To realize the loop-filter, instead of a conventional differential amplifier, we propose to use a differential difference amplifier (DDA) that is adapted to additionally serve the purpose of an IA by providing balanced high input impedance for the fully differential input signal, and lower input referred noise. Implementation of other AFE functions like programmable amplification, chopping, and filtering have also been incorporated. The resulting compact front-end architecture is suitable for power and cost efficient, high resolution, dc-to-medium speed applications.

The authors are with the Department of Electronics & Electrical Communication Engineering, Indian Institute of Technology Kharagpur, Kharagpur, Pin 721302, India (e-mail: joydeepkgp@yahoo.com).

xxxxxxxxxxxxxxxxxxxxxxxxxxx.

xxxxxxxxxxxxxxxxxxxxxxxxxxxxxxxxxxxxxxxxxxxxxxxxxxxxxxxxxxxxxx xxxxxxxxxxxxxxxxxxxxxxxxxxxx.

The rest of this paper is organized as follows. In Section II, ΔΣM and perspective of its usage in signal acquisition systems have been discussed. In Section III, the proposed front-end architecture has been presented. Design of constituent circuit blocks of the proposed system have been discussed in Section IV. Simulation and measured results along with comparison with the reported literature have been provided in Section V. Finally, conclusion has been drawn in Section VI.

## II. Background

The presented signal acquisition AFE is based on the ΔΣM topology. So, a brief overview of the standard architecture and operation of ΔΣM is pertinent. In addition, the prospect of its application in signal acquisition has also been discussed here.

### *A. The ΔΣ Modulator/ADC*

A ΔΣM performs *oversampling* as the sampling rate ($f_s$) is higher than the corresponding Nyquist rate by a factor of *OSR* (oversampling ratio >>1) [4]–[7]. This reduces the amount of quantization noise within signal band. Further, *quantization noise shaping* also occurs, that pushes most of the noise out of passband. Then, the decimation filter removes noise out of the band of interest leading to significant enhancement of signal-to-noise ratio (SNR). It also reduces output data rate (towards Nyquist rate) and increases the word length (resolution) of the ADC output. A typical discrete-time (DT) ΔΣM (as in Fig. 2) consists of a DT loop-filter *H(z)* having low-pass filtering nature (frequently a switched-capacitor (SC) integrator) and a quantizer present in a negative feedback loop. The feedback forces the average of the digital-to-analog converter (DAC) output to be equal to the input signal. As the DAC output is nothing but analog version of what the ΔΣM produces, thus, the modulator output is a digital bit stream whose average is an approximation of the input. Different ΔΣM architectures have been reported for performance enhancement, including higher order modulators where a larger fraction of the total quantization noise power is removed out of the signal band, but might have stability-related problems [5], [6].

Continuous-time (CT) version of ΔΣM has a CT loop-filter *H(s)* (usually having R-C or gm-C integrator), and the sampler is relocated before the quantizer [5], [8]. In comparison to DT counterpart, benefits of CT ΔΣM include a relaxed unity-gain frequency (UGF) requirement for constituent amplifiers or equivalently higher operating speed, lower power requirement, better noise immunity due to inherent anti-alias filtering, absence of *kT/C* sampling noise, and reduction of the noise generated from sampling process due to its shaping; and is suitable for the reduced supply voltages required for very deep sub-micron processes. In contrast, DT ΔΣM achieves better linearity, tolerance to clock jitter, and higher accuracy of the filter transfer function that depends on capacitor ratios and scales proportionally with the applied clock frequency. ΔΣ ADCs with a hybrid continuous-discrete-time approach have also been reported that attempts to utilize the advantages of both the versions [6], [9]. Such a hybrid topology has been utilized in the present work as well.

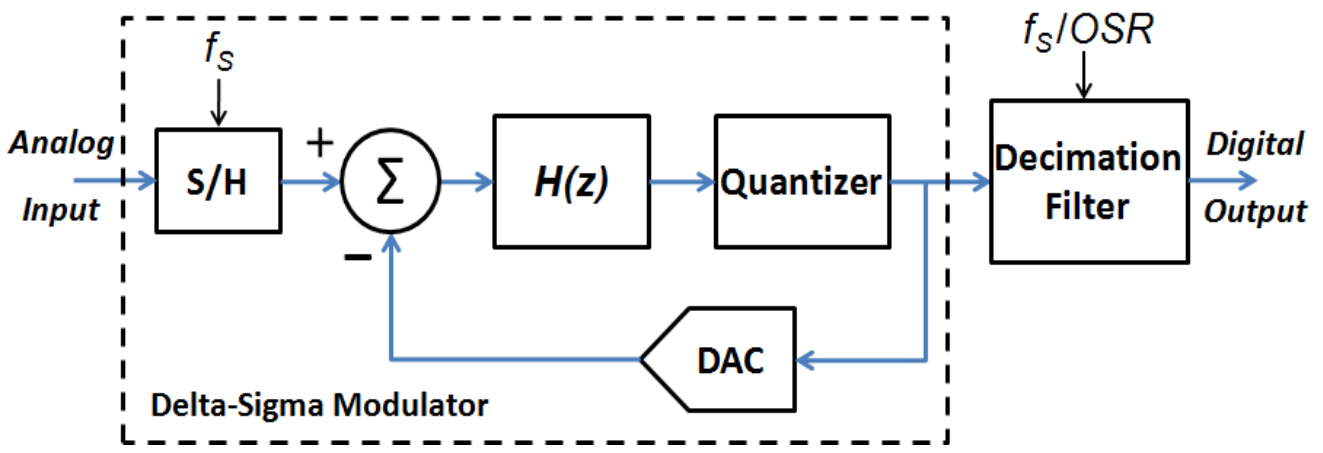

Fig. 2. Block diagram of first-order discrete-time ΔΣ ADC.

### *B. ΔΣ ADC in Signal Acquisition Front-end*

ΔΣ ADCs are utilized in the measurement of temperature, pressure, strain and vibration, bio-signal acquisition, audio recording, touch-screen sensing, battery management, and such other instrumentation that involves the measurement of low frequency signals (dc to a few 10s of kHz) with a high resolution of 12–18 bits in general [4], [10], [11]. ΔΣ ADC competes with *successive approximation register* (SAR) ADC that operates at a similar frequency range consuming quite lower power, but yields comparatively inferior resolution (8–10 bits) [12]–[14]. However, the sensor signal is required to be amplified significantly before passing to ADC for achieving the necessary resolution, as SAR ADCs typically operate at substantially high signal levels. A high-pass filter (HPF) is also required to remove dc input component so that it does not clip or saturate the IA, and hence, the ADC [14], [16]. As time constant required of the HPF is quite high, the filter consumes large area. Besides, Nyquist rate sampling imposes stringent AAF requirements. So, considerable amount of analog signal processing gets involved before digitization of the signal using SAR ADC, the penalty for which is often not included in the ADC energy metrics [17]. In contrast, due to wider dynamic range (DR) of ΔΣ ADC, it is possible to process the sensor signal with much lower amplification so that even presence of large dc-level in the signal would not drive the ADC into saturation. Hence, extra circuit blocks like the HPF, additional gain stage, and the steep AAF can be removed providing an overall reduction of complexity, size, and cost [15]. Also, dc information is not lost and can be used if needed. Moreover, ΔΣ has higher tolerance to circuit non-idealities, component mismatch errors, and is highly linear by virtue of the negative feedback. Also, presence of the digital filter makes it highly flexible towards additional post-processing requirements.

## III. System Description

The proposed compact signal acquisition AFE architecture is depicted in Fig. 3. A fully-differential (FD) second-order single-loop 1-bit ΔΣM design [7] has been used for simplicity. Two integrators have been used in cascade to constitute the loop-filter, followed by a 1-bit quantizer. A CT integrator has been used in the first stage paired with a DT integrator in the second stage. A single-bit DAC controlled by the state of the modulator digital output ($Q_p$, $Q_n$) provides the feedback signal to both the integrator stages. Using this topology, the ΔΣM has been adapted to incorporate all functionalities related to signal

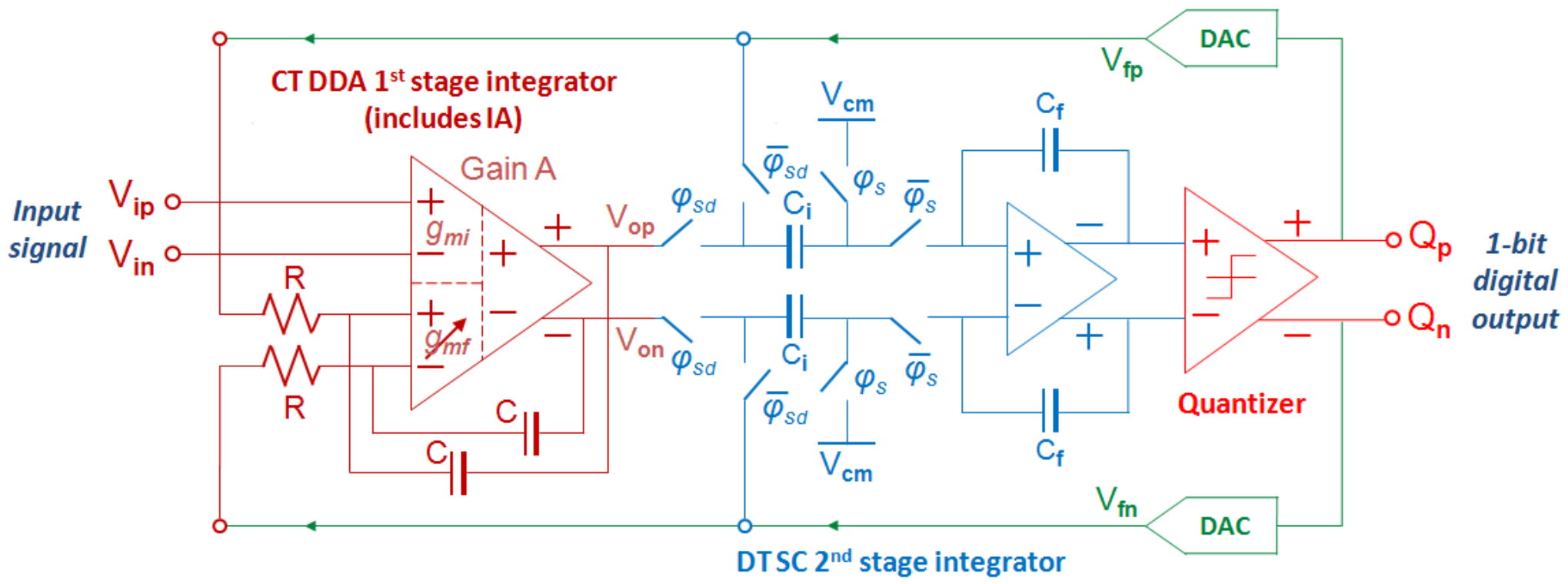

Fig. 3. Schematic diagram of proposed second-order ΔΣ modulator based compact signal acquisition front-end.

acquisition, as will be described in the following sub-sections.

*A. Combined ΔΣ-Loop-filter and IA*

In contrast to conventional ΔΣM topologies utilizing normal differential amplifier (opamp), we propose to implement the first integrator by a DDA based FD CT R-C integrator (as depicted in Fig. 3) [18] having the following transfer function.

$$V_o = (V_{op} - V_{on}) = \left\{ \frac{A(1+sRC)}{1+(A+1)sRC} V_i - \frac{A}{1+(A+1)sRC} V_f \right\} \quad (1)$$

where, $V_i = (V_{ip} - V_{in})$ and $V_f = (V_{fp} - V_{fn})$

Here, $A$ is voltage gain of DDA (for both of its non-inverting and inverting input ports). An illustrative magnitude frequency response of the integrator (along with open-loop response of constituent opamp) is provided in Fig. 4 (assuming integrator UGF of $5\times10^4$ rad/s, opamp dc gain of 60 dB, and opamp UGF of 100 kHz.). Thus, in DDA R-C integrator, both signal paths from differential input ports $V_i$ and $V_f$ to the output $V_o$ have poles at $1/(A+1)RC$ rad/s similar to that in normal opamp based R-C integrator. As $A$ is large yet finite, quantization noise shaping can get degraded due to this pole. But, the effect is minimal if $A \approx OSR$ or larger [6]. Also, transfer function from $V_i$ to $V_o$ exhibits an extra zero at $1/RC$ rad/s. As this coincides with integrator UGF and is located at much higher frequency than input signal, the desired integrator transfer function (i.e., –20dB/decade slope) is obtained for signals fed to either of the two input ports albeit with a difference in polarity.

The described DDA based integrator is used to afford the functions of both IA and ΔΣ loop-filter. That is, LPF function of IA and integrator functionality required for ΔΣM have been incorporated into the first stage of the proposed system. Note that input signal ($V_{ip}$, $V_{in}$) from sensor is directly fed to non-inverting input pair of the DDA, while the R-C feedback-network (for integrator functionality) is connected at inverting input terminals. The inverting integrator action at the port ($V_{fp}$, $V_{fn}$) is conveniently utilized to act as the feeding point for the DAC analog output required for modulator negative feedback. Thus, the input signal source remains isolated from the R-C network (in contrast to reported R-C integrator based CT

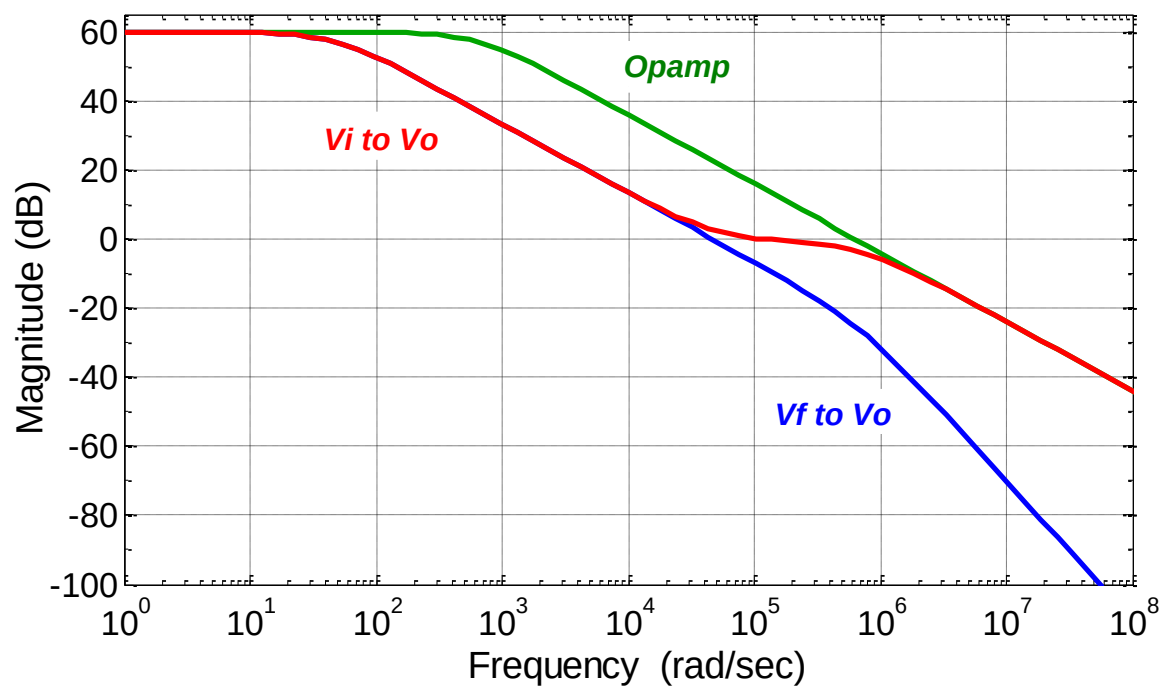

Fig. 4. Frequency response of DDA R-C integrator and constituent opamp.

ΔΣMs [6], [19]) ensuring balanced high impedance for the FD input, which is an important requisite of IA. Moreover, the proposed circuit avoids the need of a pair of large valued resistors having chip area and thermal noise implications [8]. Further, as the input common-mode (CM) voltage of the DDA can include either of the supply rails, the requirement of wide input CM range of an IA is also satisfied. Since the first stage of a signal-chain contributes most to the total noise of the system, emphasis must be given to reduce its noise. Chopping modulation might be incorporated in the DDA-based first integrator particularly for acquiring signals at low frequencies. This shifts the amplifier-induced Flicker (1/f) noise and dc-offset out of the signal band, which subsequently get removed by low-pass filtering action of the later stages [14].

Thermal noise from input resistors of R-C integrator is also a major contributor to overall noise. This has been reduced by employing an intentional difference between transconductance ($g_m$) of the input pairs at the non-inverting port ($g_{mi}$) and the inverting port ($g_{mf}$) of the DDA (as depicted in Fig. 3). By having a lower $g_{mf}$ than $g_{mi}$, the input referred noise due to the input resistors gets scaled down by the ratio of these $g_m$s (i.e., $g_{mi}/g_{mf}$). Input-referred noise of the feedback DAC also gets reduced due to same reason. This proposed scheme is unique in contrast to the recently reported design [20] that also adapts a DDA for use in a ΔΣM, some other distinguishing features being the usage of choppers and voltage feedback DAC.

A switched-capacitor DT integrator has been used as the second integrator that samples the first integrator output using clock $\varphi_s$ at rate $f_s$. Both the CT and DT integrators have been preceded by gain ($G$) of 0.5 to avoid possibility of overloading [7]. Also, UGFs of the integrators have been set to be equal at

$$\omega_{UGF,CT} = \frac{1}{RC} \equiv \frac{G}{T_s} = \omega_{UGF,DT} \tag{2}$$

with $T_s=1/f_s$ being the time-period of the oversampling clock.

Fig. 5 shows an equivalent representation of the proposed system. The effect of the high frequency zero from port $V_i$ to port $V_o$ of the DDA R-C integrator has been neglected for simplicity. The quantization error $E$ has been modeled by an additive white-noise source, while $G$ is the integrator scaling coefficient. The variable feedback voltage has been modeled by the factor $N$, while the factor $n$ models the lowered $g_m$ of the inverting input pair of the DDA.

### *B. Programmable Gain Amplification*

In the proposed topology, the feature of adjustable gain is obtained by appropriately adjusting the voltage levels ($V_{fp}$, $V_{fn}$) of the feedback signal in tune with the input signal level from the sensor. That is, for low-level input to the modulator, the feedback voltages (to both the stages) can be suitably scaled down thereby maintaining the SNR level [21]. So, a desired signal gain of $N$ is equivalent to an attenuation factor $1/N$ of the feedback DAC voltage as depicted in Fig. 5. From this model, and considering a DT equivalent for the CT first stage integrator for simplicity (and $n$=1), the modulator input-output transfer function can be expressed as follows.

$$Q = \frac{G^2 N z^{-2} V_i}{(G^2 - G + N)z^{-2} + (G - 2N)z^{-1} + N} + \frac{(1-z^{-1})^2 NE}{(G^2 - G + N)z^{-2} + (G - 2N)z^{-1} + N} \tag{3}$$

The terms with $V_i$ and $E$ denote the corresponding signal and quantization noise constituents at the output (their ratio being the signal-to-quantization-noise ratio). Thus, when feedback is attenuated (i.e. $N$>1, for low magnitude input signals), $V_i$ undergoes amplification by $N$ keeping the SNR unchanged.

For very low input levels, it becomes practically difficult to reliably generate the matching low feedback voltage. Instead of reducing the feedback, the proposed scheme of lowering the DDA gain for only the input pair in the feedback signal path (by lowering $g_{mf}$) helps. This attenuation is indicated by the factor $1/n$ in Fig. 5. As the sampling rate is much higher than the signal frequency, so, the approximation $z=exp(j\omega T_s)\approx 1$ can be used yielding the transfer function as in (4). So, the net programmable amplification available is $nN$.

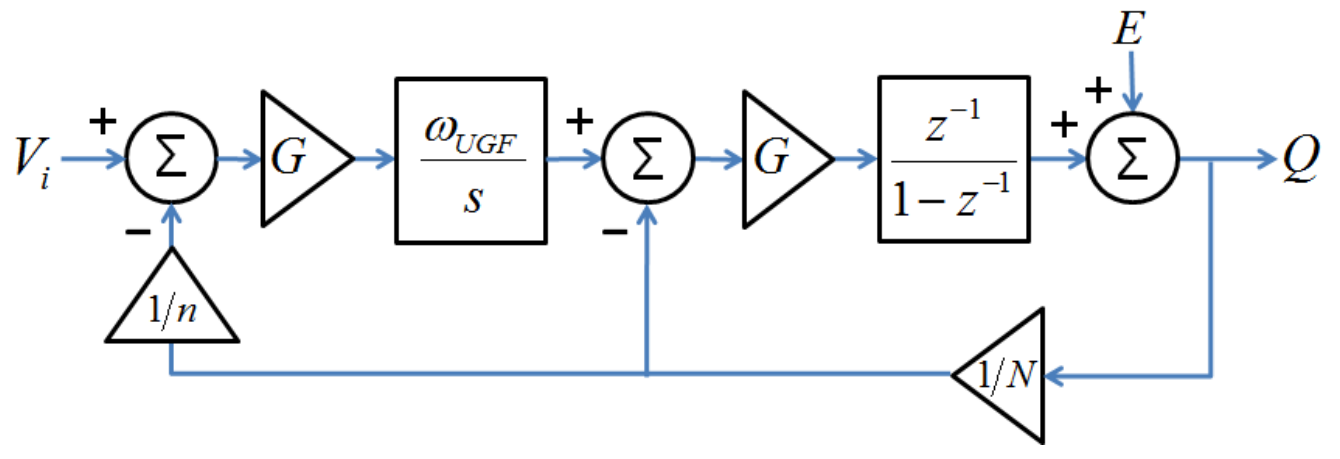

Fig. 5. Equivalent block diagram of the proposed ΔΣ modulator.

$$Q = nNz^{-2}V_i + \frac{(1-z^{-1})^2 nN}{G^2} E \tag{4}$$

### *C. Low-pass Filtering*

The low-pass nature of the CT first stage integrator enables inherent anti-aliasing property by attenuating out-of-band high frequency interferers before sampling occurs at the input of the DT second stage. In addition, as the signal is sampled at a frequency much higher than its bandwidth, so, the requirement of an explicit AAF preceeding this system becomes redundant.

A chopper-spike LPF after the anti-chopper at the DDA output is also not necessary as the clocking scheme has been implemented in a manner such that the anti-chopped signal gets sampled by the subsequent DT stage at the middle of the chopping clock phase, when the signals have settled after the chopping activity, thus avoiding sampling of voltage spikes.

### *D. Additional Aspects*

The proposed system produces single-bit digital data that represents the input signal. A multi-bit implementation of the quantizer can also be employed for enhancing SNR by virtue of its lower in-band quantization noise, reduced sensitivity to clock jitter, and improved loop stability allowing aggressive noise-shaping in higher order designs. But, the nonlinearity of associated multi-bit DAC entails additional circuitry like dynamic element matching logic and the like [8]. Clock jitter (and also, the excess loop delay) issue of CT ΔΣMs is not a concern in the low frequency of operation, and also because the proposed system is a hybrid and lower order ΔΣ [6]. So, a single-bit quantizer has been used in this design for its high linearity along with reduced complexity, chip area, and power consumption. Besides, single-bit quantizer and feedback DAC also facilitate porting of the design to lower supply voltages, hence, yielding even better power efficiency.

The ΔΣM can be coupled with standard decimation filter to realize ADC. As realization of such a filter in digital domain is quite straightforward, so the present work does not focus on design of the same. Further, instead of SC integrator in the second stage, a CT integrator can also be used in the presented architecture. Also, proposed DDA-based integrator can as well be used for its discussed benefits in other ΔΣM topologies, like higher-order/multi-bit, single-loop and cascade (or multi-loop) systems for achieving higher dynamic range.

## IV. Circuit Design and Simulation

The proposed readout front-end has been implemented at transistor level in 0.18-$\mu$m CMOS technology for evaluating the performance. The supply voltage is 1.8 V. The objective is to acquire signals in dc to 1 kHz range achieving a peak SNR of 80 dB, while limiting the power consumption to 100 $\mu$W. Main constituent blocks of the system and related design challenges have been discussed in the following sub-sections.

### *A. CT First Stage*

The first integrator is DDA based active R-C integrator with chopping modulation, as in Fig. 6(a). The DDA has been designed using FD folded cascode (FDFC) topology, as shown in the schematic in Fig. 6(b). As the opamp in the first stage has highest performance requirement, it consumes major part of the modulator power. It is also the circuit noise from ΔΣM first stage, namely, thermal noise of integrator input resistors and feedback DAC, and noise of the DDA that dominates. It is because this noise does not get shaped by ΔΣ loop unlike noise from subsequent stages. Hence, it defines the input-referred noise, and consequently, the achievable SNR of the modulator.

The UGF of the opamp of CT first stage is required to be of the order of $f_s$ [6], [19]. As too high gain is unnecessary, so, only a single stage has been used in the amplifier for power efficiency, and for ease of ensuring stability of the CM and differential feedback loops around the opamp. The input differential pairs ($M_1$–$M_4$) being one of the primary noise contributors, PMOS has been favored over NMOS for their realization in order to lower the 1/f noise. The noise of the bias transistors ($M_5$, $M_6$, $M_{11}$ and $M_{12}$) can be reduced by reducing their $g_m$ within required overdrive voltage limit. In contrast, cascode devices ($M_7$–$M_{10}$) produce much less input-referred noise due to source degeneration. Moderate to weak inversion (or, subthreshold) region of operation of the input transistors has been utilized for greater transconductance efficiency ($g_m/I_D$) [14], [19]. This is well suited for the low-power and modest bandwidth requirements of present acquisition system.

An option to reduce the DDA gain ($A$) for its second input pair (that receives the feedback signal) is required to lower the input-referred noise of the resistors of the R-C integrator and the feedback DAC, and also for the programmable gain action, as discussed earlier. This has been realized as shown in Fig. 6(b), through suitable reduction of corresponding bias current by means of turning the PMOS switch $M_{16}$ OFF. The switch must have a big enough aspect-ratio for lower resistance, so as to minimize the current mismatch between the two input pairs of the DDA. Source degeneration of this input pair can also be utilized alternatively (or, in combination with the current tuning) in order to achieve the gain reduction. Although the input range of the open-loop transconductor $g_{mi}$ is limited because of its linearity, nonetheless, this is not an issue as the expected input level is also low for the present application. Another aspect is the probable input CM mismatch between the DDA input pairs $g_{mi}$ and $g_{mf}$. However, it has been verified from simulation that there is no appreciable effect on the system performance for a mismatch of up to ±10% between the respective CM levels.

Differing from the usual CT opamp usage, a SC common-mode-feedback (CMFB) circuit has been used for stabilizing the DDA output CM as seen in Fig. 6(c). This is better than a CT CMFB in terms of linearity, power consumption, and loop stability; and helps to achieve much higher DDA output swing such that higher levels of input signals can be processed as well. Relative sizing of the switched capacitors ($C_S$) and the non-switched CM-sensing capacitors ($C_C$) involves a trade-off [22]–[24]. Having $C_S$ bigger (~5x) than $C_C$ results in quicker settling of the output CM after start-up and lower steady-state errors. But, the settling time at each clock cycle will be more. In contrast, a smaller $C_S$ not only reduces capacitive loading of the opamp, but, also lowers the equivalent SC resistive loading that tends to degrade the opamp dc-gain [22]. Preventing the latter is important as any reduction in dc-gain of constituent amplifier causes shifting of the integrator pole away from dc, deteriorating the modulator's noise shaping performance. The capacitor values chosen here are $C_S$=20 fF and $C_C$=100 fF. The value of $C_S$ being somewhat small might get affected due to parasitic capacitance and process variations. However, it has been verified from simulation that the effect of any resultant skew between the two switched capacitors on the overall system performance is negligible. The switches have been realized by CMOS transmission gate (TG) using minimum sized transistors for diminishing the effect of charge injection on produced feedback voltage ($V_{fb}$). The complementary non-overlapping clocks ($\varphi_s$ & $\overline{\varphi_s}$) used in subsequent DT second

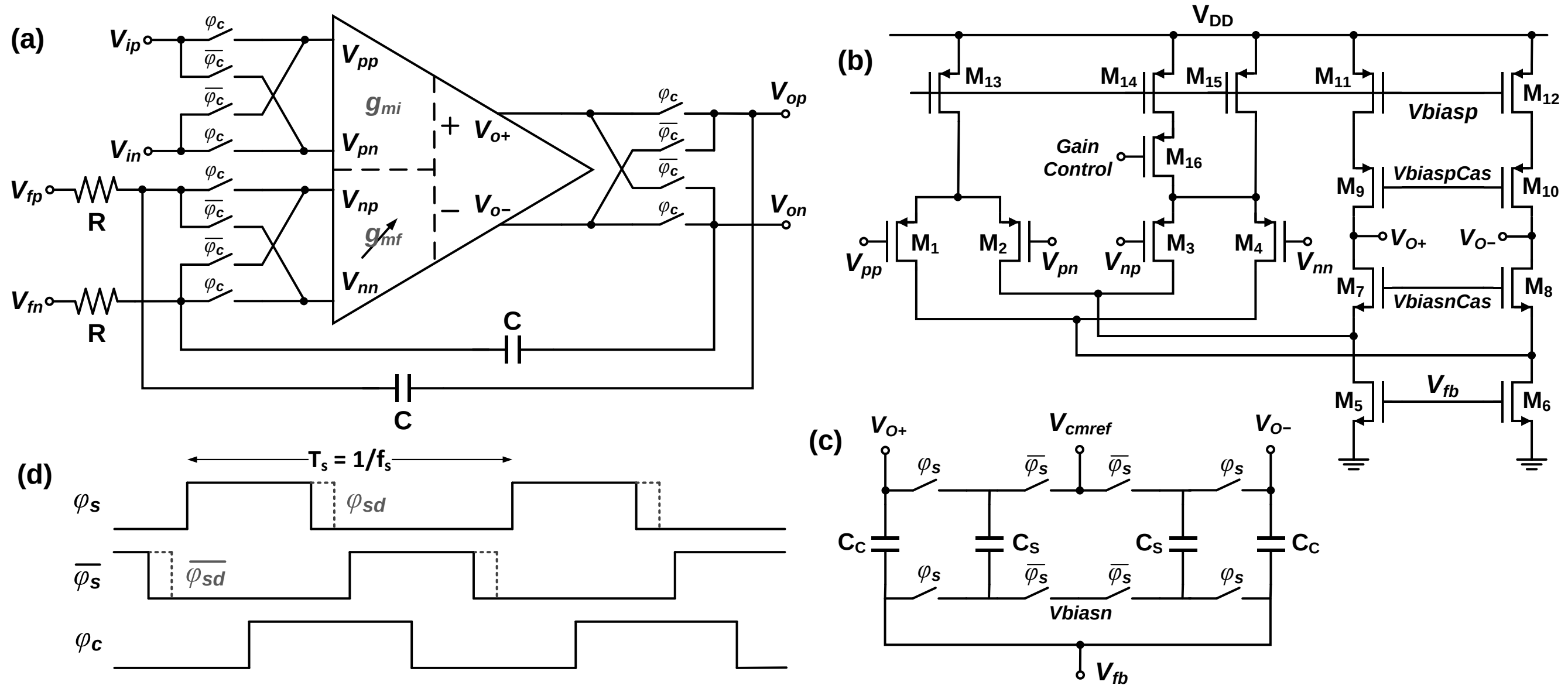

Fig. 6. CT first stage: (a) chopped DDA based R-C integrator, (b) DDA circuit topology with gain-control for inverting input differential pair, (c) switched capacitor CMFB circuit used to stabilize the output common-mode voltage of the DDA, and (d) relation between the sampling and chopping clock phases.

integrator have been used to control switches of this CMFB.

The first integrator has been chopper stabilized by using crossed switches that periodically invert the signals at the input and output ports of the DDA, as depicted in Fig. 6(a). Chopping in ΔΣM is typically done at half the oversampling rate as it yields better noise removal compared to a lower chopper frequency (such as $f_s/4$, $f_s/8$,…) [25], [26]. However, the output spectrum of a modulator contains large amount of noise at out-of-band frequencies, particularly at $f_s/2$. Due to any parasitic coupling of the $f_s/2$ chopping clock into the ΔΣM output or feedback branches, down modulation of the high frequency noise existing at $f_s/2$ harmonics into the baseband might occur. This noise in turn affects at the feedback input of the modulator, thereby degrading the SNR [27], [28]. To avoid this in the present system, chopping has been done at $f_{ch}=f_s$. Further, chopping operation is performed at the middle of the sampling clock phases as illustrated in Fig. 6(d). This provides enough time for the DDA to settle from disturbances (at both the input and output ports) as caused by chopping, and thereby prevents sampling of chopper spikes by next integrator stage. The chopper switches have been realized with CMOS TGs carefully sized to avoid non-linearities caused by them.

The AFE in this work (or, the first stage) has been designed to have a dc-coupling, so that it can process both ac and dc input signals. But, if required for a particular application, the topology can be augmented to ac-coupled using techniques as in [16], [29]. In general, ac-coupled IA is favoured for bio signals that can have large differential dc offset. In contrast, sensors like strain gauge, accelerometer, thermocouple, Hall sensor, resistance temperature detector, etc. can produce dc output as well and does not have dc offset voltage.

*B. Device Noise Analysis*

Voltage noise power spectral density (PSD) due to device noise sources at the DDA R-C integrator output is given by

$$S_{noise,out}(f)=\frac{8kTRA^2}{1+(2\pi f(A+1)RC)^2}+\left\{S_{DDA,th}\left(1+\frac{f_c}{f}\right)\frac{A^2[1+(2\pi fRC)^2]}{[1+(2\pi f(A+1)RC)^2]}\right\} \tag{5}$$

Here, $S_{DDA,th}$ represents the thermal noise PSD of the DDA referred to its non-inverting input, and $f_c$ represents the corner frequency at which the 1/f noise spectrum equals its thermal noise PSD. Thus, at very low frequencies, Flicker noise of the DDA dominates, while the spectrum becomes white towards high frequencies. Now, if the integrator is chopper modulated, the resultant PSD at output node can be derived as follows. Here, only the chopper modulated noise component at the first harmonic at $f=f_{ch}$ has been considered.

$$S_{noise,out,ch}(f)=\frac{8kTRA^2}{1+(2\pi f(A+1)RC)^2}+\frac{8}{\pi^2}S_{DDA,th}\times\left(1+\frac{f_c}{|f-f_{ch}|}\right)\frac{A^2[1+(2\pi(f-f_{ch})RC)^2]}{[1+(2\pi(f-f_{ch})(A+1)RC)^2]} \tag{6}$$

Noise simulation results for the integrator presented in Fig. 7 correspond well with these theoretically derived expressions. Using the output PSD, the input-referred noise PSD of the chopped DDA based integrator can be derived as follows

$$S_{noise,in,ch}(f)=\frac{8kTR}{1+(2\pi fRC)^2}+\frac{8}{\pi^2}\left(1+\frac{f_c}{|f-f_{ch}|}\right)\times S_{DDA,th}\frac{[1+(2\pi(f-f_{ch})RC)^2][(1+(2\pi f(A+1)RC)^2]}{[1+(2\pi(f-f_{ch})(A+1)RC)^2][1+(2\pi fRC)^2]} \tag{7}$$

Thus, it is obvious that thermal noise of the resistors ($R$) and DDA are required to be decreased in order to reduce noise of the chopped integrator. Thermal noise from the input resistors being a dominant noise source, determines the upper limit of R that can be used [6]. However, the proposed usage of different $g_m$ for the two input pairs of the DDA leads to different gains from the two input pairs to the DDA output. Say, $A_i$ and $A_f$ are the gains from the non-inverting input and the inverting input respectively to the output; with $A_i$ being greater than $A_f$. Thus, the input referred noise due to the resistors gets scaled down by the ratio of the associated gains, consequently reducing the output noise as evident from the corresponding noise PSD provided below (and also from the plot in Fig. 7).

$$S_{noise,out}(f)=\frac{8kTRA_f^2}{1+(2\pi f(A_f+1)RC)^2}+\left\{S_{DDA,th}\left(1+\frac{f_c}{f}\right)\frac{A_i^2[1+(2\pi fRC)^2]}{[1+(2\pi f(A_f+1)RC)^2]}\right\} \tag{8}$$

Noise generated by the inverting input pair transistors also gets scaled down by the same factor when referred to the non-inverting input port, consequently, reducing $S_{DDA,th}$ as well. Hence, a substantial lowering of the device noise of the first stage integrator can be achieved by means of using the DDA based R-C integrator with skewed transconductances for its two input pairs. Additionally, when chopping is enabled, the device noise PSD at the output becomes as follows.

$$S_{noise,out,ch}(f)=\frac{8kTRA_f^2}{1+(2\pi f(A_f+1)RC)^2}+\frac{8}{\pi^2}S_{DDA,th}\times\left(1+\frac{f_c}{|f-f_{ch}|}\right)\frac{A_i^2[1+(2\pi(f-f_{ch})RC)^2]}{[1+(2\pi(f-f_{ch})(A_f+1)RC)^2]} \tag{9}$$

Here also, the chopper modulated DDA noise component at the first harmonic (at $f=f_{ch}$) has only been considered for being the dominant term in comparison to all other harmonics. Also, the noise spectrum referred to the DDA non-inverting port is

$$S_{noise,in,ch}(f)=\frac{8kTRA_f^2}{A_i^2[1+(2\pi fRC)^2]}+\frac{8}{\pi^2}\left(1+\frac{f_c}{|f-f_{ch}|}\right)\times S_{DDA,th}\frac{[1+(2\pi(f-f_{ch})RC)^2][(1+(2\pi f(A_f+1)RC)^2]}{[1+(2\pi(f-f_{ch})(A_f+1)RC)^2][1+(2\pi fRC)^2]} \tag{10}$$

Thus, the input referred noise from the resistors indeed get scaled down by the ratio of the gains $A_f$ to $A_i$. It should be noted that increase in noise at higher frequencies (as in Fig. 7) due to the different $g_m$ technique will not affect the system performance as the input signal would always lie in the lower frequency zone where gain of the constituent opamp is high.

### C. Integrator R-C Selection

Selection of suitable values of R and C for first integrator is important due to impact on the overall power, noise, and area of the system. A sampling frequency of 1 MHz has been chosen yielding an *OSR* of 500. Thus, from (2), $RC=2T_s=2\ \mu$s for which $R$=100 kΩ and $C$=20 pF have been used. Although for second order ΔΣM, maximum OSR of about 256 is enough as any further increase in $f_s$ does not reduce in-band noise due to dominance of circuit thermal noise over quantization noise [30]. Nonetheless, a somewhat large $f_s$ has been deliberately chosen in order to reduce the *RC* value, hence requiring less layout areas for the on-chip resistors and capacitors. Smaller valued resistor also produces lower thermal noise, and smaller load capacitor helps in achieving higher UGF of DDA. The capacitors have been realized using Metal-Insulator-Metal (MIM) capacitor for good voltage linearity and matching. The minimum DDA UGF required is of the order of $f_s$ [4], [19]. So, the $g_m$ required, and consequently the sizes of the input transistors are also large. Moreover, a large $g_{mi}$ helps to reduce the input referred noise, and larger transistor sizes reduce the Flicker noise and mismatch between the input devices.

### D. DT Second Stage

The second integrator is a SC integrator having parasitic insensitive topology [6], realized using FDFC opamp with SC CMFB. Although the opamp UGF required for realizing a DT ΔΣM is higher (for proper settling accuracy) than that for a CT version, the opamp requirements get relaxed beyond the first stage integrator due to the attenuation of corresponding noise and distortion by the preceding integrator(s) via noise shaping [5]. As with the first stage DDA, moderate inversion operation of the input differential pair has been used for higher energy and noise efficiency. The sizing of the CMFB capacitors and switches, and the clocking utilized for it are the same as that for the SC CMFB of the DDA as illustrated in Fig. 6(c).

The gain $G$=0.5 at input of the integrator has been obtained by selecting capacitor values such that $C_i/C_f$=0.5. The $kT/C$ thermal noise produced by sampling switch on-resistances is inversely proportional to sampling capacitor ($C_i$) value. But, large capacitances increase loading on opamp and fabrication area, leading to a trade-off in setting the capacitor values. As the noise contribution of the second stage integrator is small compared to first stage, hence, capacitors of the order of 1 pF are adequate. Due to larger parasitic capacitance to substrate from bottom-plate of MIM capacitors, opamp input terminals have been connected to the top-plate to reduce substrate noise getting coupled to the sensitive input; whereas, the bottom-plate is driven either by the input voltage source or the opamp output. To control the sampling and integration operations of the integrator, non-overlapped clock phases $\varphi_s$ and $\overline{\varphi}_s$ along with corresponding delayed versions $\varphi_{sd}$ and $\overline{\varphi}_{sd}$ have been used, as shown in Fig. 6(d). Delaying of only the falling edges is sufficient for minimizing error due to signal dependent channel charge injection from the switches [6], [7]. To prevent harmonic distortion, linear TG switches for the integrator have been realized with transistors sized to keep the on-resistance approximately constant over the input voltage range.

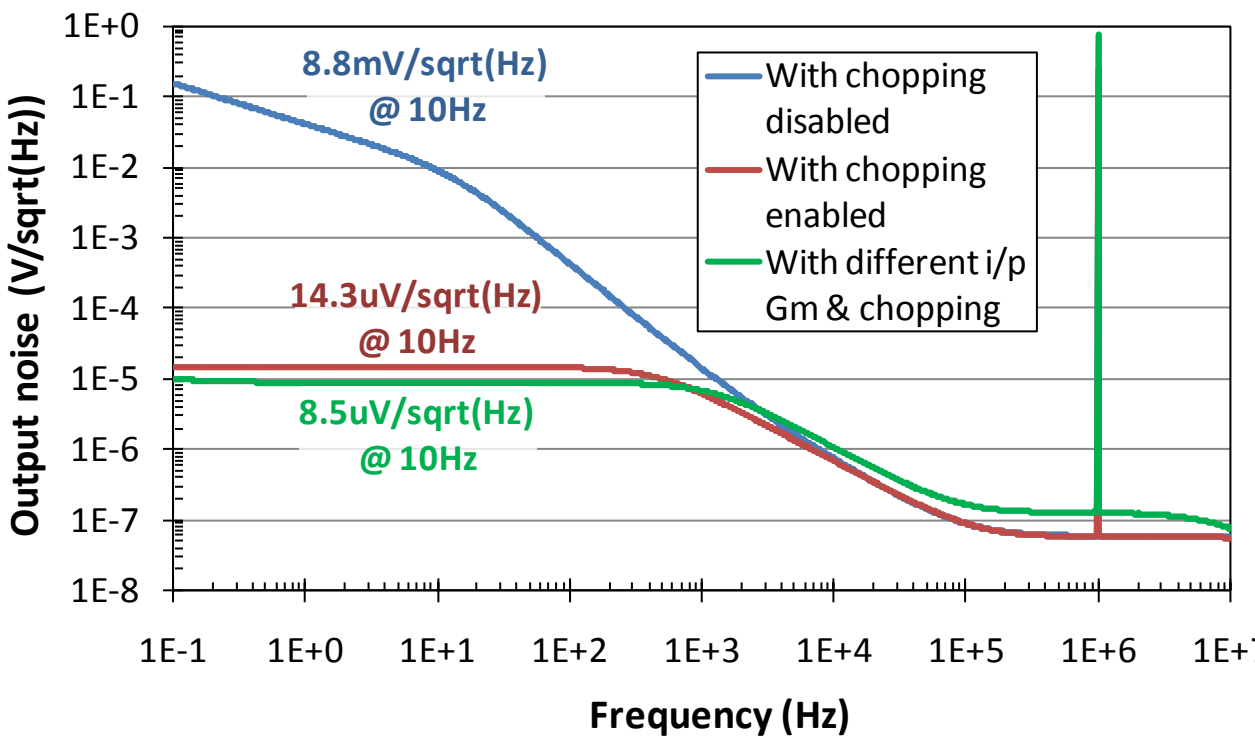

Fig. 7. DDA integrator simulated output device noise PSD.

### E. Current Mirror Bias Circuit

Bias voltages for amplifiers in the system have been derived using self-biased wide-swing cascode *Beta-multiplier* current mirror circuit [23] as in Fig. 8. It is combination of NMOS wide-swing cascode current mirror comprising of transistors $M_1$–$M_4$ and an equivalent PMOS current mirror comprising $M_8$–$M_{11}$, connected as a positive feedback loop. The diode-connected $M_5$ generates the voltage *VbiasnCas* providing bias to the NMOS cascode transistors $M_1$ and $M_4$. Current for the biasing transistor is derived from the same bias-loop through $M_6$ and $M_7$. Diode connected $M_1$ and $M_2$ receives the current from output of the upper PMOS current mirror, and mirrors it to produce output current from $M_4$ (and also yields *Vbiasn*). The PMOS half of the mirror operates likewise. Resistance $R_\beta$ and ratio of the sizes of $M_2$ and $M_3$ can be tuned to set the current magnitude. The generated current is independent of supply-voltage ($V_{DD}$) variations, and has good temperature behavior as well. Filtering capacitors ($C_F$) ~ 1 pF have been utilized between the generated voltages and the supply rails to filter any high frequency noise.

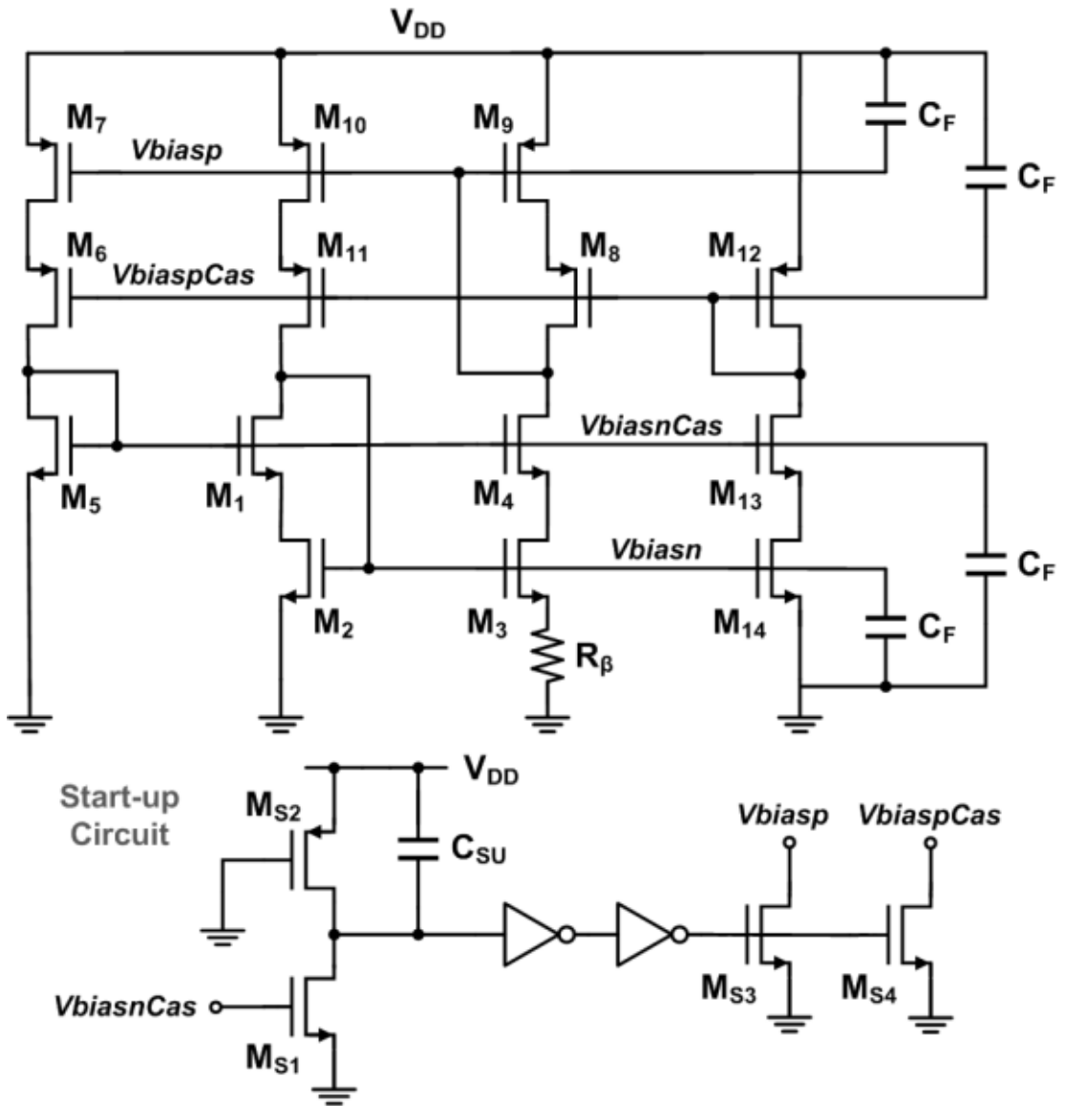

Fig. 8. Wide-swing cascode Beta-multiplier current reference circuit.

To prevent the possibility of the circuit being stuck at the unwanted stable operating point having zero current in all the transistors, a *start-up circuit* has been used as shown in Fig. 8. If this state occurs, then $M_{S2}$ pulls the gates of $M_{S3}$ and $M_{S4}$ towards $V_{DD}$, thereby pulling gates of the PMOS transistors low. This starts flow of current in the circuit and it eventually settles to the desired state. Once in normal region of operation (i.e., non-zero quiescent current), $M_{S1}$ (which is sized much stronger than $M_{S2}$) becomes ON and turns OFF $M_{S3}$ and $M_{S4}$, thus, disabling the start-up circuit. A relatively big capacitor $C_{SU}$ ~ 5 pF has been added to extend the start-up duration, thus preventing oscillation in the start-up feedback loop.

### *F. Quantizer*

A regenerative comparator comprising of a dynamic latch followed by a static NOR based SR-latch has been utilized for realizing the 1-bit quantizer [5]–[7], [31], as seen in Fig. 9. The comparator is strobed by SC integrator sampling-phase clock $\varphi_s$ so as to generate a decision at the beginning of this phase when the preceding integrator output remains stable. The regeneration phase starts when $\varphi_s$ goes high. The dynamic positive-feedback due to cross-coupled inverters ($M_{1A}$/$M_{2A}$ & $M_{1B}$/$M_{2B}$) swiftly amplify the differential input voltage ($V_{inp}$, $V_{inn}$) into full-scale decision (output) voltage that is buffered and fed to the digital SR-latch. The current path between the supplies gets broken after a decision has been made. When the clock $\varphi_s$ goes low, $M_{5A}$ & $M_{5B}$ turn ON causing the latch to reset, and both of its outputs ($V_{op}$, $V_{on}$) are pulled low, and hence, the SR-latch preserves the state saved in the previous regeneration phase. Simultaneously, $M_4$ turns OFF and snaps the positive feedback of the latch, and also the current path between supplies, thereby avoiding static power consumption. The devices $M_{1A}$ & $M_{1B}$ also provide adequate isolation of the input terminals from kickback disturbances from regeneration nodes [31] at the beginning of reset and regeneration phases.

Size of the input pair ($M_{3A}$ & $M_{3B}$) has been kept large enough (hence, having high $g_m$) to produce adequate current for quick and proper regeneration, and reduce process induced mismatch. The switches ($M_{5A}$, $M_{5B}$ and $M_4$) are with minimum length to decrease their on-resistance and charge injection. To prevent '11' input to SR-latch during the brief metastable state of the regenerative latch, PMOS transistors in the cross-coupled inverters need to be suitably larger than NMOS so that their threshold voltage (or, metastable level) becomes higher than the input threshold of the succeeding stage. Moderate sizing of the cross-coupled inverters ensures optimum capacitance at the nodes $V_{oA}$ and $V_{oB}$ for quick regeneration and reduced process sensitivity. The buffering inverters should be sized small to minimize effect of output-state dependent capacitive loading (of regeneration nodes) due to following SR-latch, that might affect the regeneration for low levels of differential input. Nonetheless, performance requirements of the quantizer is quite relaxed for single-bit ΔΣMs, as the non-idealities of this stage get largely suppressed by the preceding integrator stages in a manner similar to the quantization error. A pre-amplifier may be additionally used before the dynamic latch to improve the comparator resolution while reducing the offset and kick-back from the regenerative latch. But, even without it, the topology is capable of achieving the required resolution, hence saving power [32].

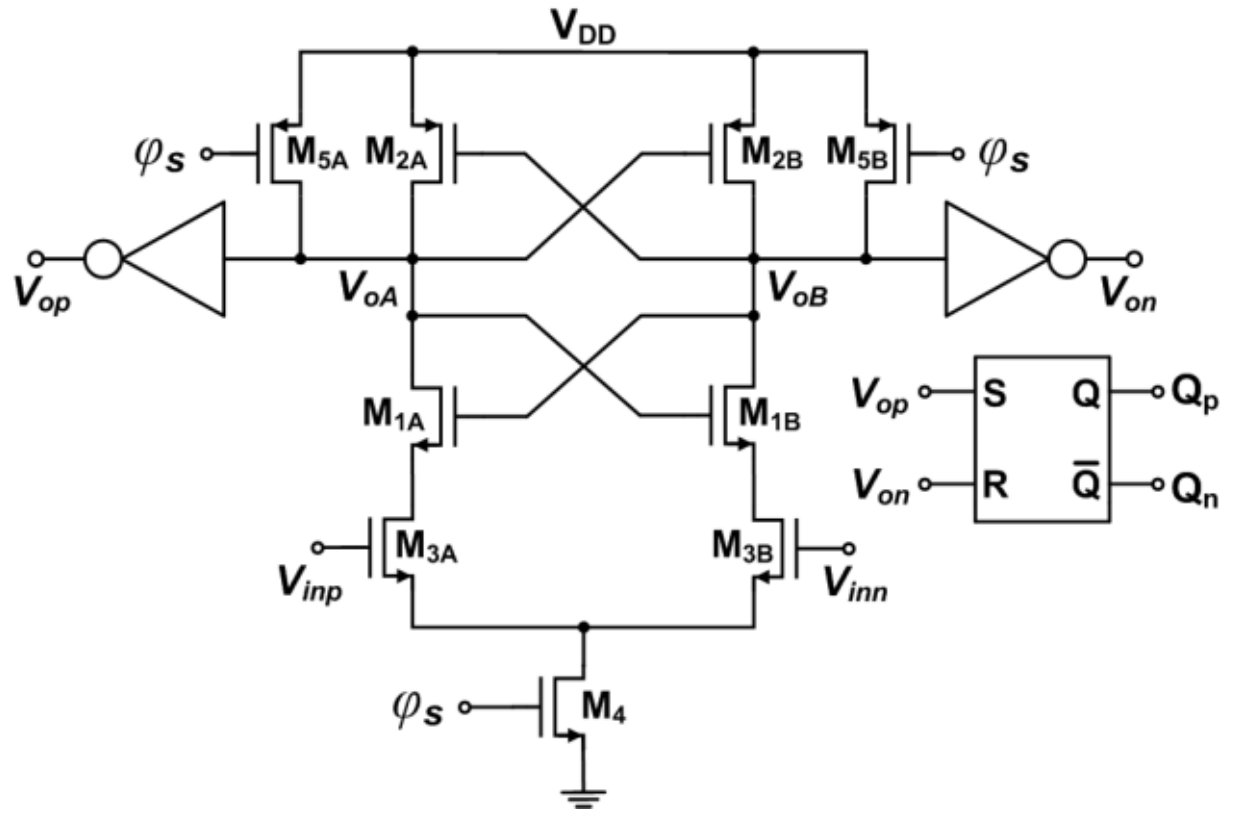

Fig. 9. Circuit schematic of the regenerative feedback comparator.

### *G. Clock Generation*

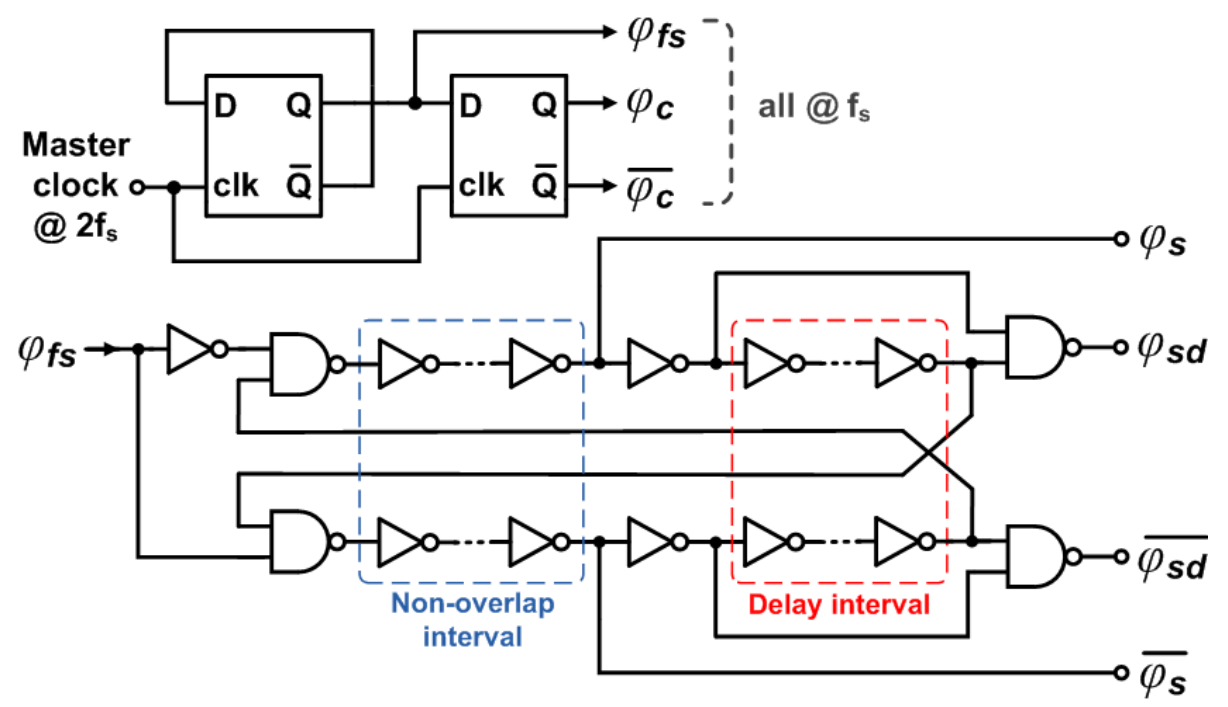

Fig. 10. Clock phase generator circuit schematic.

Non-overlapping clocks that are never high at the same time and their delayed versions required for the operation of SC integrator and CMFB circuits, and the clocks required for the choppers have been generated on-chip using the circuit shown in Fig. 10. A complement version of each clock phase is also required for the TGs. The circuit receives external clock signal (at twice the required $f_s$) and creates clock $\varphi_{fs}$ at half the input rate and also π/2 phase-shifted complementary versions ($\varphi_c$ and $\bar{\varphi}_c$) required for chopping. Complementary versions of $\varphi_{fs}$ are provided to a pair of NANDs gates that drive two delay lines interconnected by feedback paths, producing the non-overlapped phases and their delayed versions [5], [6]. The first set in the cascade of inverters has an odd number of inverters and controls the non-overlap period; while the delaying of phases is produced by passing through the second set having an even number of inverters. Transistor aspect ratio and the number of inverters in these two sets have been designed in order to achieve desired non-overlap and delay periods. The second pair of NANDs ensure that only falling edges of the clocks get delayed. The nominal delay and non-overlapping times are 2.5 ns and 3 ns respectively, while the rise and fall times are around 40 ps. The capacitive load that each clock phase signal has to drive might differ significantly leading to a detrimental influence on the delays and non-overlap times. Hence, all generated clock phases have been buffered to get required fan-out before being fed to the application circuits.

### *H. Reference Voltage Generation & DAC*

The reference dc voltage levels $V_{refp}$ and $V_{refn}$ required for feedback DAC as well as the CM voltage $V_{cm}$ (=$V_{DD}$/2) have been derived using the circuitry shown in Fig. 11(a). Multiple options of the reference levels (e.g., $V_{cm}$ ± 20/50/100.. mV) are generated via a series of resistors between supply and ground, as required for the programmable gain feature which demands variable feedback levels. Depending on magnitude of the input signal, a particular set of reference voltages is required to be manually selected from the available options using appropriate MOS switches, and then buffered before being fed to ΔΣM. Adaptive tuning can also be appended so that the voltages automatically get adjusted according to the input strength.

The value of resistances in the ladder have been optimized to lower dc current while keeping their thermal noise in check (as the voltages are used for feedback DAC). Further, bypass capacitors ($C_F$) have been added to the reference lines, that also reduce the effect of noise by restricting noise bandwidth. The output impedance of the buffers have been adequately lowered in order to avoid significant drops in the reference voltages [6], [33]. A FC opamp cascaded with an appropriate source-follower stage (as illustrated in Fig. 11(b)), in unity negative feedback has been used to realize the buffers of Fig. 11(a). The resultant output impedance is the small source-follower output impedance divided by the large loop-gain, which turns out to be of the order of a few ohms. Depending on whether the reference voltage level to be buffered is above or below $V_{cm}$, either PMOS or an NMOS source-follower stage has been employed to reduce systematic output offset in the buffered voltages. Regarding the buffer used for $V_{cm}$, an extra level-shifter stage has been added as in Fig. 11(b). Since the buffer loop-bandwidth is finite, its output impedance tends to increase at higher frequencies. Also, the generated voltages being used in SC circuitry, incur transient glitches and noise due to high frequency switching. To minimize such effects, large off-chip bypass capacitors ($C_{ext}$) ~ 20 $\mu$F have been used. This also lowers the output impedance at high frequencies, and helps to compensate the negative feedback loop of the buffers as well. However, parasitic inductance of the package bond-wires (required for connecting the external capacitors) forms an under-damped parallel resonant circuit with on-chip capacitances [34]. Thus, whenever there is a current variation in the bond-wires due to switching in the SC circuits, the reference voltages tend to ring. To diminish such oscillations, parallel on-chip R-C damping networks as shown in Fig. 11(c) have been utilized [33]. MOS capacitor has been used in the damping circuit for $V_{cm}$ for saving area as high linearity is not a requisite. However, as the difference between $V_{refp}$ and $V_{refn}$ may not be adequate to bias a MOS in strong-inversion, MIM capacitor has been used between those two references. The values of $R_D$ and $C_D$ have been suitably set using simulation for minimizing the settling time of ringing.

The feedback non-return-to-zero (NRZ) DACs of Fig. 11(d) have been implemented with CMOS switches that select either of the two reference voltage levels ($V_{refp}$ and $V_{refn}$), and are controlled by the single-bit digital output ($Q_p$ and $Q_n$) of the modulator. Thermal noise from the DAC is a dominant source and should be lowered by using sufficiently large switches.

## V. System Performance and Comparison

Operation of proposed system architecture has been verified by behavioral modeling in *MATLAB Simulink*. Subsequently, circuit level simulation has been executed in *Cadence Spectre* for verification of performance. Further, parasitics extracted netlist from full-chip layout as designed for the proposed AFE has been simulated, and the results have been found to match well with that from corresponding schematic simulation. In this regard, it should be mentioned that standard physical design techniques have been adopted in preparing the layouts of the constituent blocks. This includes appropriate separation between the analog, digital, and mixed-signal constituent blocks (to protect sensitive analog from aggressor digital parts) and their power supplies, common-centroid layout and dummy placement for matched devices, having adequate number of n-well and p-substrate contacts in the form of rings to prevent the possibility of latch-up, symmetrical routing of differential signal traces, usage of bypass capacitors between $V_{DD}$ and Gnd, etc. Thereafter, the design has been fabricated in SCL's (an Indian foundry **Error! Reference source not found.**) 0.18-$\mu$m CMOS process and the resulting test-chip has been characterized. Fig. 12 shows the corresponding die image and measurement setup.

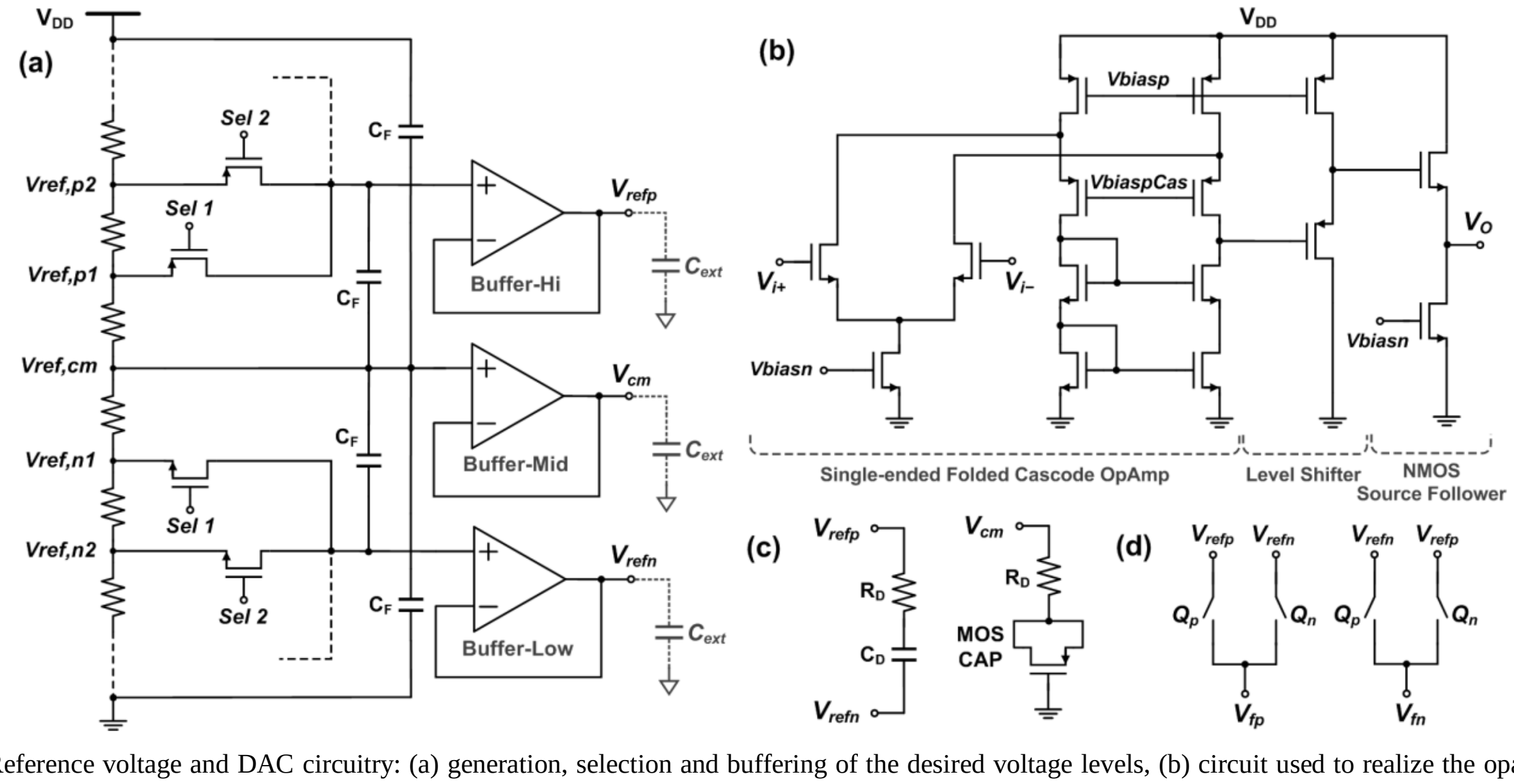

Fig. 11. Reference voltage and DAC circuitry: (a) generation, selection and buffering of the desired voltage levels, (b) circuit used to realize the opamp of the Buffer-Mid analog-buffer, (c) RC damping networks, and (d) feedback DAC realization using CMOS switches.

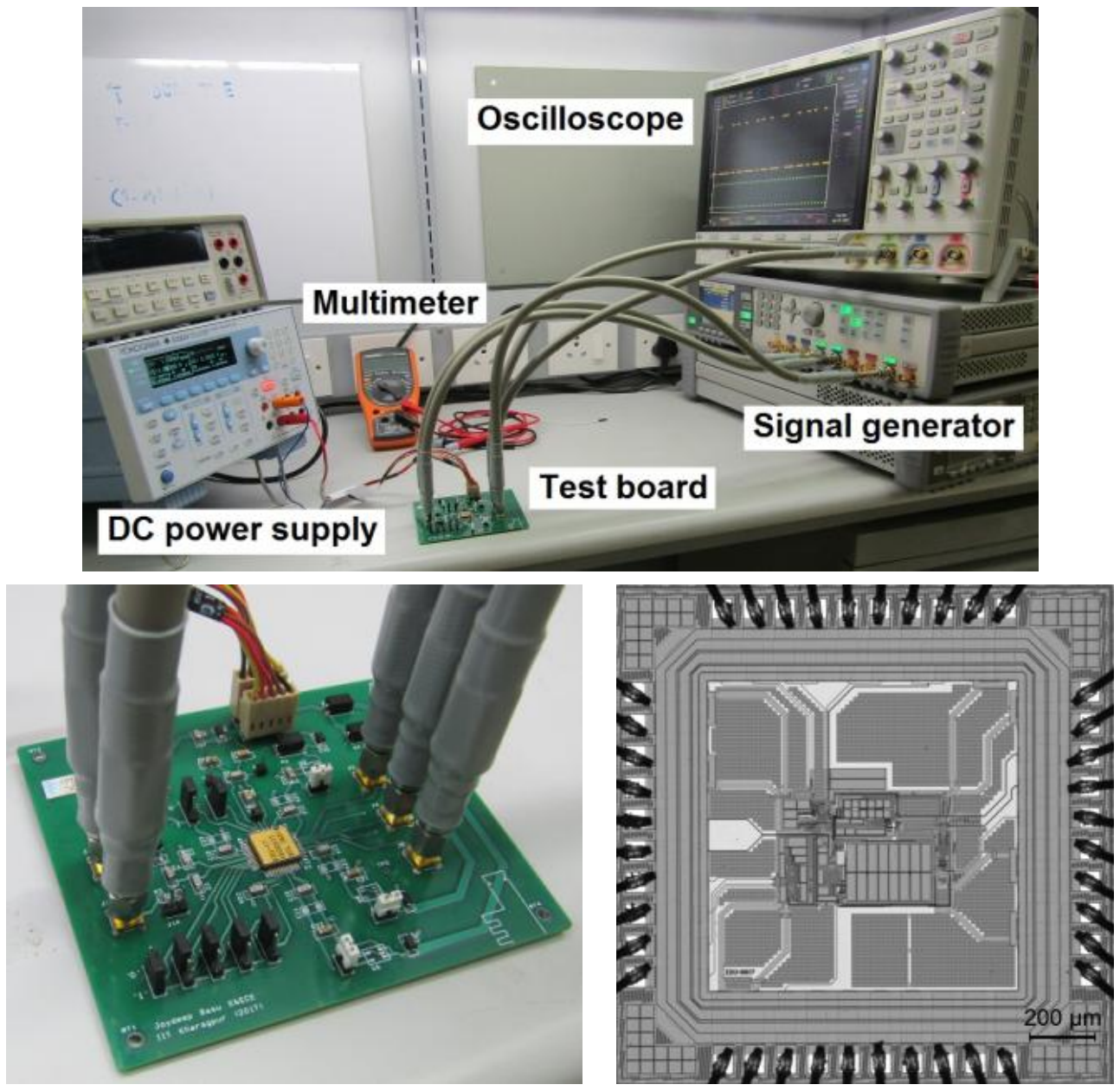

Fig. 12. Picture of the test setup along with test board and fabricated die.

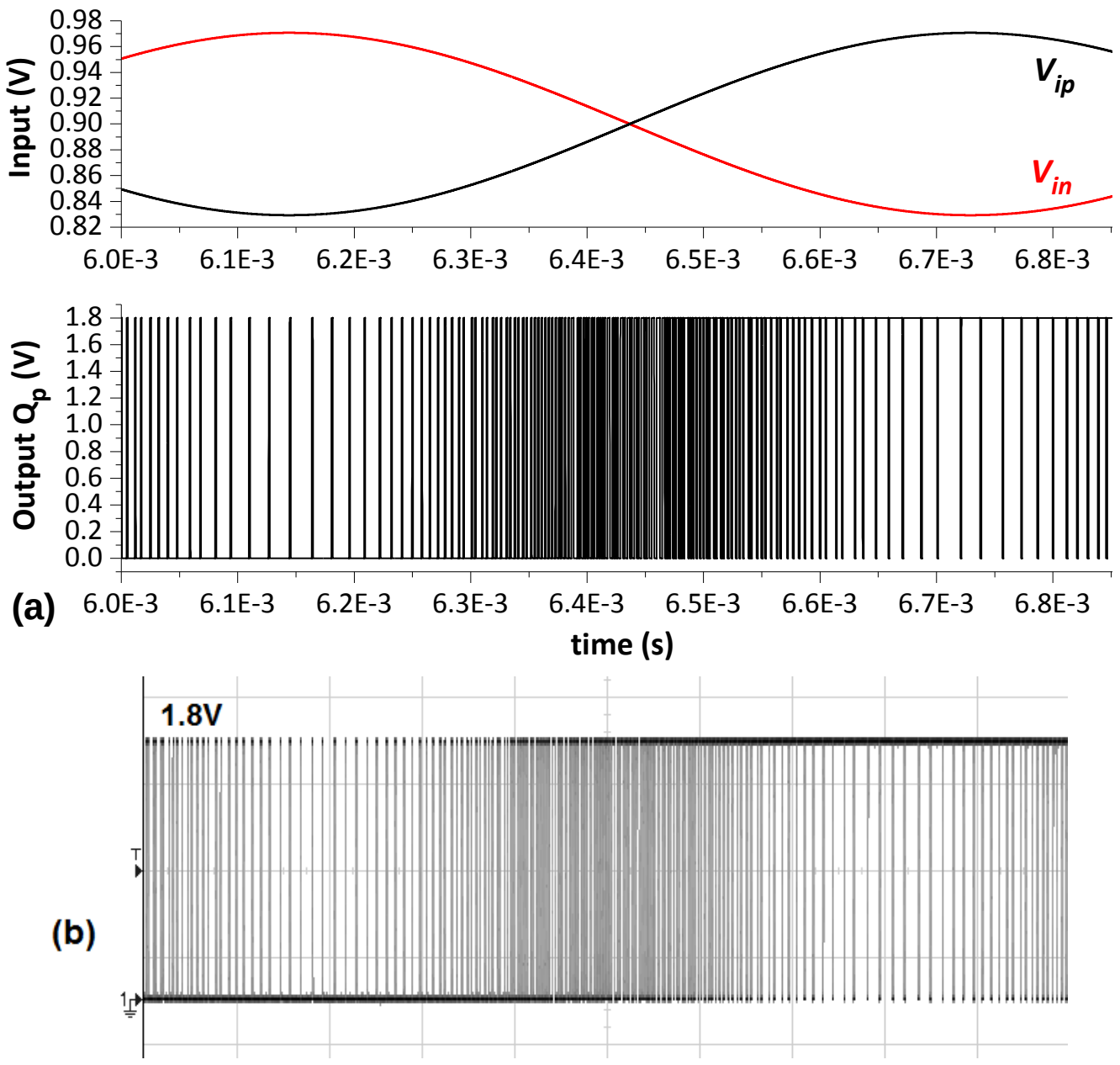

Fig. 13. Transient input and output (both simulated (a) and measured (b)) waveforms of the proposed system.

Transient simulation result of the designed ΔΣM is provided in Fig. 13(a). This shows a FD sine input ($V_{ip}$, $V_{in}$) of 854.5 Hz as provided to the system and the 1-bit digital output ($Q_p$). Fig. 13(b) shows the corresponding output as obtained from the test-chip, having the expected pulse-density modulated nature with relatively more number of '1' bits when the input is high, more number of '0' bits when the input is low, and almost equal number of '1's and '0's when the signal is in mid range. In order to find the system's output SNR in presence of device noise sources, *transient noise* simulation [36] has been utilized in Spectre. Discrete Fourier Transform (DFT) has then been performed on the transient output for analyzing frequency components in the signal. The resultant 8192-point DFT plot utilizing *Hanning window* is presented in Fig. 14. It is evident that the fabricated chip exhibits a higher level of device noise than that predicted by simulation.

An outline of the simulated performance of overall system is presented in Table I for an input signal amplitude of 71 mV at typical PVT (process, voltage and temperature) corner. Results both without and with device noise sources considered have been presented. The performance of the system has also been checked at a lower $V_{DD}$ of 1.2 V with correspondingly scaled input amplitude of 47 mV, the results of which are also provided in Table I. This proves the operational viability of the system at lower supply voltage, hence, yielding higher

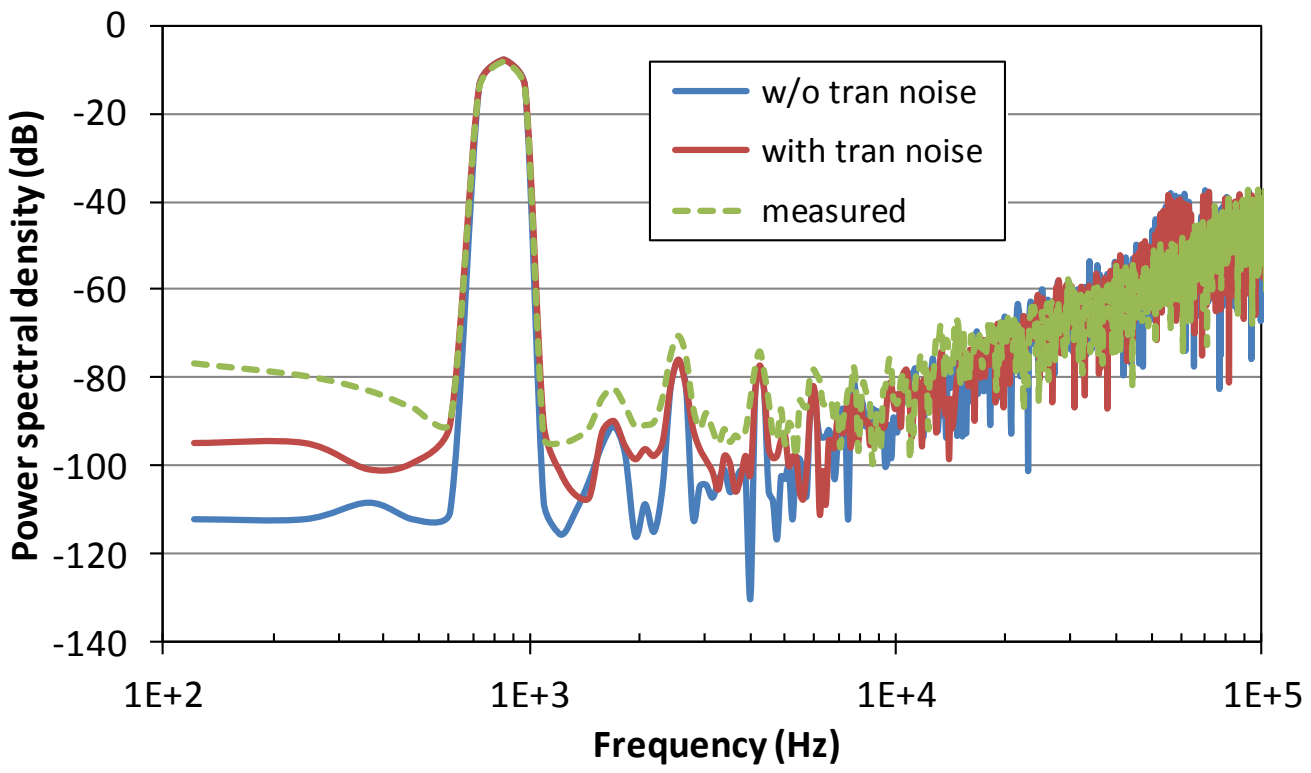

Fig. 14. Power spectral density (DFT) of the modulator output signal.

TABLE I
COMPLETE SYSTEM SIMULATED PERFORMANCE

| Parameter | Specification | Simulation (w/o noise) | Simulation (with noise) | Simulation (with noise)* |
|---|---|---|---|---|
| Input Signal Bandwidth | 1kHz | | | |
| Input Signal Frequency | ~ 1kHz | 854.5Hz | | |
| Sampling Frequency | 1MHz | | | |
| OSR | ~ 500 | | | |
| SNR | 74–86dB | 92.6dB | 80.1dB | 80.5dB |
| ENOB | 12–14 bits | 15.1 bits | 13.0 bits | 13.1 bits |
| Power | ~ 100µW | 98.4µW | 98.4µW | 66.4µW |

* This is using $V_{DD}$ of 1.2V; rest all data are at $V_{DD}$ of 1.8V

power efficiency. The performance of the system has also been checked at all the significant PVT corners as in Table II, with nominal variation being observed across the different corners. Further, *Monte Carlo* mismatch analyses (with 100 samples) have been performed on the constituent blocks as well as on the complete modulator. The results of the latter are provided in Fig. 15 (here device noise sources have not been enabled to save simulation time) showing only minor variation in performance due to mismatches.

The variation of SNR with input signal amplitude from 0dBFS to –80dBFS is provided in Fig. 16. In this plot, 0dBFS refers to a maximum input signal amplitude of 100 mV. The simulated DR of the present system is more than 80 dB (for maximum input level of 100 mV, without considering device noise sources to save simulation time). As the minimum test signal amplitude available from the utilized signal generator was only 25 mV, the measured DR plot could not be obtained at lower input amplitude levels.

It must be emphasized that the system can process inputs over a much wider amplitude range due to the provision of variable feedback voltages, thereby, enabling PGA action. The corresponding simulation results (including device noise) are provided in Table III. Thus, peak SNR has been obtained at a minimum input amplitude of about 70.71 mV (at feedback of 100 mV). Also, it was noted that the performance tends to get limited by third harmonic distortion at higher input magnitude to the DDA. The designed AFE can detect from a minimum input level of 0.001 mV to a maximum level of 300 mV, hence, yielding an overall DR of 109.54dBFS with respect to full-scale amplitude of 300 mV. However, this value could not be measured from test-chip due to limitation of instrument.

A comparative analysis of the measured (and post-layout simulated) performance of the designed system with reported performance of similar ΔΣMs (all having comparable order and operating frequency) is presented in Table IV. Walden figure-of-merit ($FOM_W$) expression [4] as given in (11) has been utilized for the comparison. For ΔΣM, a smaller value of $FOM_W$ indicates better performance. Comparison has also been done using Schreier figure-of-merit ($FOM_S$) [5] as in (12), for which a larger value indicates a more efficient ADC.

$$FOM_W = \frac{Power\ consumption\ (W)}{2^{ENOB(bits)} \cdot Nyquist\ rate\ (S/s)} \times 10^{12} (pJ/conversion) \quad (11)$$

$$FOM_S = DR(dB) + 10\log\frac{Bandwidth(Hz)}{Power\ consumption\ (W)} (dB) \quad (12)$$

It may be noted that, except for the design in [42], all other designs in literature require input signal with reasonably large amplitude to provide the quoted performance. In other words, in order to use those designs in a signal acquisition system, an accompanying suitable instrumentation amplifier is needed. Whereas, the present system can process inputs of much lower amplitude in comparison to the other systems in Table IV (except for [42] which has a comparable input amplitude, but much higher power and hence, inferior $FOM_W$). Therefore, it should be highlighted that although the proposed system is a

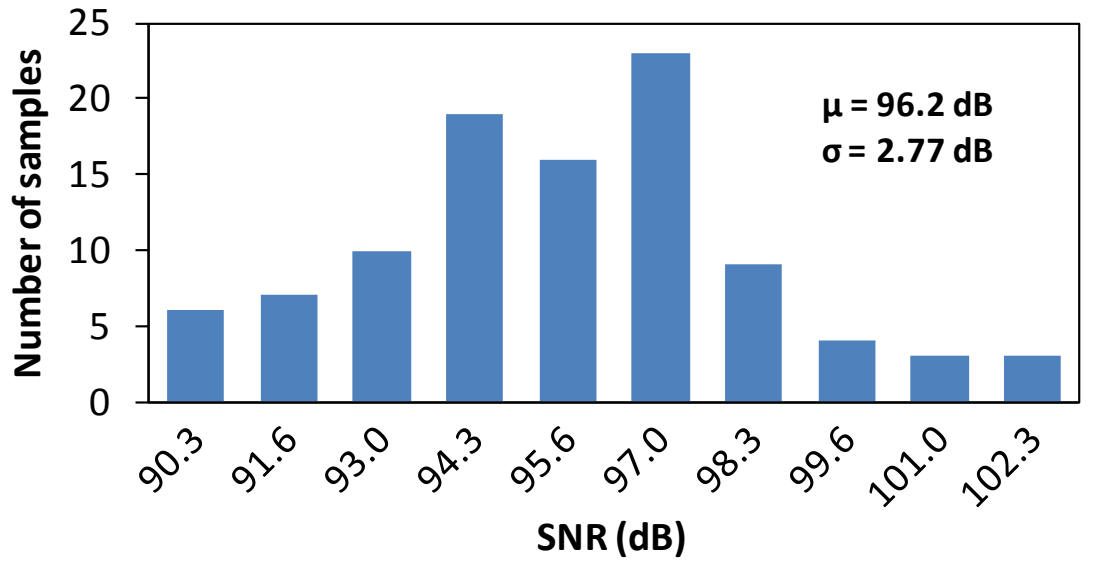

Fig. 15. SNR of the AFE as obtained from Monte Carlo simulation. Here, μ and σ denote the mean and standard deviation values respectively.

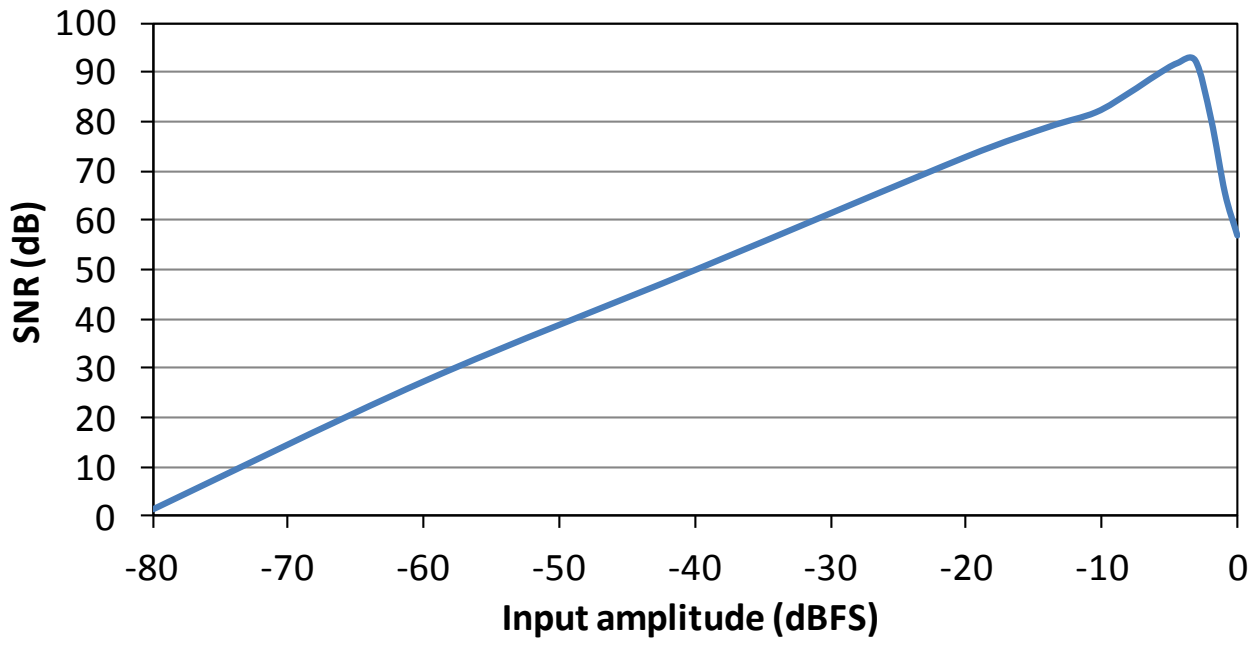

Fig. 16. Output SNR versus the input signal level.

TABLE II
COMPLETE SYSTEM PERFORMANCE ACROSS PVT CORNERS

| NMOS | PMOS | Resistor | $V_{DD}$ (V) | Temp (°C) | SNR (dB) | ENOB (bits) | Power (μW) | Comments |
|---|---|---|---|---|---|---|---|---|
| T | T | T | 1.8 | 27 | 80.1 | 13 | 98.4 | Typical corner |
| S | F | T | 1.8 | 27 | 82.7 | 13.4 | 99 | Operating pt. is okay across S-F & F-S |
| F | S | T | 1.8 | 27 | 79.4 | 12.9 | 103.8 | |
| F | F | T | 1.95 | -20 | 83.2 | 13.5 | 91 | Highest speed corner |
| S | S | T | 1.6 | 125 | 81.7 | 13.3 | 94.8 | Slowest speed corner |
| T | T | Min | 1.8 | 27 | 83.5 | 13.6 | 111.2 | Not much change due to resistance variation |
| T | T | Max | 1.8 | 27 | 80.3 | 13 | 89.5 | |

Abbreviations – T: typical, S: slow, F: fast corner

TABLE III
SNR OF DESIGNED SYSTEM AT DIFFERENT INPUT AND FEEDBACK LEVELS

| Feedback voltage (mV) (0dBFS) | Input amplitude (mV) | SNR (simulated with noise) (dB) | SNR (measured) (dB) |
|---|---|---|---|
| 20 | 0.001 (-109.5dBFS*) | 0.32 | — # |
| 20 | 14.14 (-3dBFS) | 71.3 | — # |
| 50 | 35.36 (-3dBFS) | 76.5 | 67.8 |
| 100 | 70.71 (-3dBFS) | 80.1 | 70.8 |
| 300 | 212.13 (-3dBFS) | 80.8 | 70.9 |

* This is with respect to full-scale amplitude of 300 mV
\# Could not be measured due to amplitude limitation of the test instrument

ΔΣ converter, it embodies the additional function of sensor signal conditioning within the same FOM. The reason behind emphasis on simulated results throughout the paper (and not just the measured performance) is the process dependence of the amount of device noise that in turn, affects the obtainable SNR. Thus, it is believed that the same design can yield higher measured SNR if implemented in a better foundry (as also

TABLE IV
PERFORMANCE COMPARISON OF PROPOSED AFE WITH REPORTED SIMILAR MODULATORS

| Process | BW | $f_s$ | OSR | Power | ENOB (bits) | DR (dB) | $FOM_S$ (dB) | $FOM_W$ (pJ/conv) | Differential input amplitude/range | Chip active area | Reference |
|---|---|---|---|---|---|---|---|---|---|---|---|
| CMOS 180 nm | 100 Hz | 150 kHz | 750 | 505 μW | 18 | — | 163 | 9.7 | 4.4 Vp-p | 0.8 $mm^2$ | Xu et al., 2015 [37] |
| CMOS 2000 nm | 1 kHz | 500 kHz | 250 | 940 μW | 13.8 | 84 | 144 | 32.9 | 5 Vp-p | 2 $mm^2$ | Nadeem et al., 1994 [38] |
| CMOS 130nm | 150 Hz | 2.2 MHz | 7300 | 1.44 mW | 18 | 112 | 162 | 18.3 | 3 Vp-p | 0.4 $mm^2$ | Fraisse et al., 2016 [39] |
| CMOS 180 nm | 256 Hz | 57 kHz | 111 | 13.3 μW | 11.7 | 83 | 156 | 7.8 | 1.4 Vp-p | 0.51 $mm^2$ | Cannillo et al., 2011 [40] |
| CMOS 350 nm | 1 kHz | 640 kHz | 320 | 12.7 mW | — | 136 | 185 | — | 8.6 Vp-p | 11.48 $mm^2$ | Steiner et al., 2016 [41] |
| CMOS 700 nm | 100 Hz | 300 kHz | 1500 | 2 mW | 12.8 | 82 | 129 | 1400 | 100 mVp-p | 3.3 $mm^2$ (with IO ring) | Sarhangnejad et al., 2011 [42] |
| CMOS 150 nm | 2 kHz | 320 kHz | 80 | 96 μW | 11 | 68 | 141 | 14.6 | 0.7 Vp-p | 1.02 $mm^2$ | Garcia et al., 2013 [43] |
| CMOS 180 nm | 1 kHz | 1 MHz | 500 | 99.2 μW | 13.1* 11.5 | 109* | 179* | 5.6* 17.1 | peak SNR at 141 mVp-p | 0.16 $mm^2$ | This work (at $V_{DD}$ = 1.8V) |
| CMOS 180 nm | 1 kHz | 1 MHz | 500 | 67 μW | 12.9* 11.8 | 109* | 180* | 4.4* 9.3 | peak SNR at 94 mVp-p | 0.16 $mm^2$ | This work (at $V_{DD}$ = 1.2V) |

* This is from post-layout simulation (taking device noise sources into account), rest are based on measurement

verified through comparative simulations). Nonetheless, the objective of this paper is not to outperform the state-of-the-art, but to illustrate the operational viability and performance of the proposed system architecture through accurate simulations and measurement. A further enhancement of its noise and power performance is feasible through application of various prevalent aggressive techniques for the corresponding aspects.

## VI. CONCLUSION

A wide range of applications and surging demand of various sensors make it imperative to investigate the associated signal acquisition front-end circuitry in an attempt to enhance the performance of the overall system. In view of the advantages of using ΔΣ ADC in readout front-end, this work proposes a ΔΣM based compact signal acquisition system architecture. In this architecture, a DDA R-C integrator adapted to act as both instrumentation amplifier and as integrator of ΔΣM loop has been proposed. Other signal acquisition functionalities, like balanced high input impedance and programmable gain for the input signal, low-pass filtering, and digitization have also been incorporated into the system. Further, difference between the transconductances of the DDA input pairs has been proposed to reduce the effect of input resistor thermal noise of the front-end R-C integrator. Chopper modulation has been utilized as well in order to minimize the effect of Flicker noise. Thus, the proposed architecture is an aggregation of the entire signal acquisition system functions within the single block of ΔΣM. This translates to a size, cost, and power efficient signal acquisition AFE for high precision applications, as verified through comprehensive simulation. Further, a robust operation across process corners has also been verified for the system.

The design was fabricated in SCL 0.18-$\mu$m process. Post-layout simulation as well as measured performance of the system has been presented along with a comparison of the same with that of reported similar ΔΣMs. Peak simulated SNR of 80 dB at a minimum input amplitude of 47 mV, and an overall dynamic range of about 109dBFS has been obtained for an input signal band of 1 kHz while consuming 100 $\mu$W of power. However, the measured SNR was found to be lower by about 9 dB. Improvement of noise and power aspects of the current version of AFE can be envisioned through the usage of alternative process, as well as by the application of established specialized techniques on constituent blocks of the system. The relatively low-frequency design presented here is for demonstrating the efficacy of the proposed topology, and is ideal for acquiring signal from sensors such as accelerometers, thermocouple, biosensors etc. However, it should be noted that the same signal acquisition system architecture can be adapted for higher operational frequencies as well.